\documentclass{epl}

\usepackage{amsmath,amssymb,latexsym}
\usepackage{graphicx}
\newcommand{\qed}{\hfill $\square$\\}
\newcommand{\defeq}{\stackrel{\text{def}}=}

\begin{document}

\title{Bell's inequality and the coincidence-time loophole}
\shorttitle{Bell's inequality and the coincidence-time\ldots}
  
\author{Jan-\AA ke Larsson\inst{1} \and Richard D.\ Gill\inst{2,3}}
\shortauthor{J.-\AA.\ Larsson \and R.D.\ Gill}
\institute{ \inst{1} Matematiska Institutionen, Link\"opings
  Universitet, SE-581 83 Link\"oping, Sweden.\\
  E-mail:~{\tt jalar@mai.liu.se}\\
  \inst{2} Department of Mathematics, University of Utrecht, Box
  80010, NL-3508 TA Utrecht, The Netherlands\\
  \inst{3} EURANDOM, P.O. Box 513, NL-5600 MB Eindhoven, The
  Netherlands}


\maketitle

\begin{abstract}
  This paper analyzes effects of time-dependence in the Bell
  inequality. A generalized inequality is derived for the case when
  coincidence and non-coincidence [and hence whether or not a pair
  contributes to the actual data] is controlled by timing that depends
  on the detector settings.  Needless to say, this inequality is
  violated by quantum mechanics and could be violated by experimental
  data provided that the loss of measurement pairs through failure of
  coincidence is small enough, but the quantitative bound is more
  restrictive in this case than in the previously analyzed
  ``efficiency loophole.''
\end{abstract}

The Bell inequality \cite{Bell64} and its descendants (see
\emph{e.g.}, ref.~\cite{CHSH}) have been the main argument on the
EPR-paradox \cite{EPR,BohmAhar} for the last forty years.  A new
research field of `experimental metaphysics' has formed, where the
goal is to show that the concept of local realism is inconsistent with
quantum mechanics, and ultimately with the real world. The experiments
which have been performed to verify this have not been completely
conclusive, but they point in a certain direction: Nature cannot be
described by a local realist model. The reason for saying ``not been
completely conclusive'' is the existence of certain ``loopholes'' in
these experiments. For example, the so-called ``efficiency loophole''
is present in all photonic experiments performed to date including
Aspect {\it et~al} \cite{Asp1,Asp2,Asp3} and Weihs {\it et~al}
\cite{WJSWZ}. This is usually dealt with using the ``no-enhancement''
assumption, which would not be needed if the efficiency would be high
enough (see refs \cite{Pearle70,ClauHorn,GargMerm,
  Shimony93,jalar98a,jalar99a}). Recently, there was a 100\%
efficiency experiment by Rowe {\it et~al} \cite{Rowe01}, but that
experiment does not enforce strict locality.

This paper is motivated by recent claims (\emph{e.g.},
ref.~\cite{hp-pnas1}) that time-dependence has been omitted in the
Bell inequality. A close examination of the presented
counterexample(s) show that they are in fact nonlocal; the probability
measure used depends on \emph{both} detector parameters and is thus
not local in the Bell sense. In fact, there is no problem in using the
measurement time as a hidden variable in a truly local Bell setting.
Nevertheless, while working through the details we found that timing
issues may indeed play a role, even in a local model. The present
analysis incorporates this in a generalized Bell inequality, for which
the reader may find that certain bounds are higher than na\"\i{}vely
expected. We will also note that the loophole can be closed with
relative ease in experiments, and indeed \emph{is} in some modern
experiments.

\begin{figure}[htbp]
  \begin{center}
    \includegraphics[width=.63\linewidth]{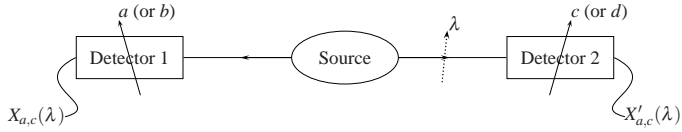}
    \caption{The Bell setup. 
      }
    \label{fig:3}
  \end{center}
\end{figure}

The situation is as follows: in the standard Bell setup (see
Fig.~\ref{fig:3}), we want to take into account that the
time-correlation is not perfect between results at one site and
results at the other. Usually, there is a ``coincidence window'' in
which events are counted as being ``simultaneous,'' i.e., belonging to
the same pair so that they contribute to the correlation, even when
there is a finite (\emph{i.e.}, nonzero) time between
them\footnote{For simplicity we assume that the two detectors are
  stationary in the same inertial frame, and use that frame to
  determine simultaneity.}. There is a possibility that, in this type
of setup, the local setting may change the time at which the local
event happens. And this would have implications; a certain local event
may be simultaneous with a remote event, or not, depending on the
local detector setting.  The result will be that the simultaneity of
two detection events will depend on \emph{both settings}, even though
the underlying physical processes that control this are completely
local. We will examine this situation in detail and derive precise
bounds for violation of the appropriate Bell inequality.

To perform the intended formal examination of this, we need to put the
hidden-variable model into formal language. We arrive at a
probabilistic model \footnote{It may seem that we only discuss the
  ``deterministic'' case here, but a generalization to the
  ``stochastic'' case is straightforward and will not be done here.}.
Here, the hidden variable is a point $\lambda$ in a ``sample space''
$\Lambda$, the space of all possible values of the hidden variable.
The measurement results are described by random variables (RVs)
$X(\lambda)$ which take their values in the value space $V$, usually
taken as $\{-1,\,+1\}$ in the spin-$\frac12$ case (see
Fig.~\ref{fig:3}).

There is a probability measure $P$ on the space $\Lambda$, used to
calculate the probabilities of the different outcomes and the
expectation value $E$, where 
\begin{equation}
  E(X)=\int_{\Lambda}X(\lambda)dP(\lambda)=\int_{\Lambda}X dP,
\end{equation}
suppressing the parentheses.
We then obtain the expectation of the product of the results as
$E(XX')$, usually denoted ``correlation'' in this context
\footnote{The reason for this is that in the simplest case
  $|X|=|X'|=1$ and $E(X)=E(X')=0$, and then the correlation is
  precisely $E(XX')$. Here, this terminology will be retained even for
  cases when this ceases to be valid, as it has become standard in
  this context}.  
Finally, after using \emph{locality} only four RVs
remain; $A$, $B$, $C'$ and $D'$, see below.
Now, we have

\smallskip\noindent \emph{Theorem 1 (The Clauser-Horne-Shimony-Holt
  (CHSH) inequality)} The following three prerequisites are assumed to
hold except at a null set:
\begin{enumerate}
\item \label{Cp1} \emph{Realism.} Measurement results can be described
  by probability theory, using two families of RVs, \emph{e.g.},
  \begin{equation}
      \begin{split}
X_{a,c}:\Lambda&\rightarrow V\\
        \lambda&\mapsto X_{a,c}(\lambda)\\
        X_{a,c}':\Lambda&\rightarrow V\\
        \lambda&\mapsto X_{a,c}'(\lambda).
      \end{split}\tag{i}
  \end{equation}
\item \label{Cp2} \emph{Locality.} A measurement result should be
  independent of the remote setting, \emph{e.g.},
  \begin{equation}
    \begin{split}
      A(\lambda)&\defeq X_{a,c}(\lambda)= X_{a,d}(\lambda)\\
      C'(\lambda)&\defeq X_{a,c}'(\lambda)=X_{b,c}'(\lambda).
    \end{split}\tag{ii}
  \end{equation}
\item \label{Cp3} \emph{Measurement result restriction.}  The results
  may only range from $-1$ to $+1$,
  \begin{equation}
    V=\{x\in{\mathbb{R}};-1\le x\le+1\}.\tag{iii}
  \end{equation}
\end{enumerate}
Then
\begin{equation}
  \big| E(AC')+E(AD') \big| + \big| E(BC')-E(BD') \big| \le 2.
\end{equation}

The proof consists of simple algebraic manipulations inside each of
the two expressions on the right hand side, followed by application of
the triangle inequality on each expression.

Previous treatments have discussed several loopholes in this
inequality, but the most similar issue to the present is the
``detector efficiency'' problem. A simple formalism to use is that of
ref.~\cite{jalar98a}, where inefficient detectors, or in more general
terms, inefficient \emph{measurement setups} are modeled by having the
measurement-result RVs \emph{undefined} at points in $\Lambda$ where
no detection occurs. This means that, \emph{e.g.}, the RVs $A$ and
$C'$ will only be defined at subsets of $\Lambda$ denoted $\Lambda_A$
and $\Lambda_{C'}$, resp.. The averaging must now be restricted to the
set where the RV in question is defined, and the probability measure
adjusted accordingly. In the language of probability theory we need
the \emph{conditional} expectation value
\begin{equation}
  E(A|\Lambda_A) = \int_{\Lambda_A} X_A dP_A,
\end{equation}
using the conditional probability measure
\begin{equation}
  P_A(S)=P(S|\Lambda_A)\text{ for all events }S.
\end{equation}
In the detector efficiency case, the first correlation in the CHSH
inequality then is $E(AC'|\Lambda_A\cap\Lambda_{C'})$, the expectation
of $AC'$ conditioned on both factors in the product being defined
(that both results are observed). This is the correlation that would
be obtained from an experimental setup where the coincidence counters
are told to ignore single particle events.

In the present case, there is a slight difference. In a
hidden-variable model, the detection times $T$ and $T'$ at the two
sites can be described as RVs that depend on the settings,
\emph{e.g.},
\begin{equation}
  \begin{split}
    T_{a,c}:\Lambda&\rightarrow \mathbb{R}\\
    \lambda&\mapsto T_{a,c}(\lambda)\\
    T_{a,c}':\Lambda&\rightarrow \mathbb{R}\\
    \lambda&\mapsto T_{a,c}'(\lambda).
  \end{split}
\end{equation}
A ``coincidence'' then occurs when the two times differ by less than
some predetermined time interval $\Delta T$. In mathematical language
this corresponds to saying that coincidences occur for certain values
of the hidden variable $\lambda$, \emph{e.g.}, at the settings $a$ and
$c$, values in the set
\begin{equation}
  \label{eq:6}
  \Lambda_{AC'}
  \defeq\Big\{\lambda:\big|T_{a,c}(\lambda)-T_{a,c}'(\lambda)\big|
  <\Delta T\Big\}.
\end{equation}
In the following, we will concentrate on the set $\Lambda_{AC'}$ (but
it does help to remember its origin), and this set can vary depending
on both detector settings. Note that no assumption has yet been made
on the locality of the detection times $T$ and $T'$ at the two sites;
they may depend on both settings.  Remember that we are concentrating
on the resulting statistics of $A$, $B$, $C'$, and $D'$ here, and only
use the times to find coincidences. If it makes the reader feel
better, he/she may use an implicit locality assumption ($T_a=T_{a,c}$
and $T'_c=T'_{a,c}$), but that is of no consequence below.

The first correlation in the CHSH inequality then is
$E(AC'|\Lambda_{AC'})$, the expectation of $AC'$ conditioned on
coincidences for the settings $a$ and $c$.  The original CHSH
inequality is no longer valid, and the reason can be seen in the start
of the proof where one wants to add
\begin{equation}
    \Big|E(AC'|\Lambda_{AC'})+E(AD'|\Lambda_{AD'})\Big|
    =\bigg|\int_{\Lambda_{AC'}} AC'dP_{AC'}+\int_{\Lambda_{AD'}}
    AD'dP_{AD'}\bigg|.
\end{equation}
The integrals on the right-hand side cannot easily be added when
$\Lambda_{AC'}\neq\Lambda_{AD'}$, since we are taking expectations over
different ensembles $\Lambda_{AC'}$ and $\Lambda_{AD'}$, with respect to
different probability measures. 

The problem here is that the ensemble on which the correlations are
evaluated changes with the settings, while the original Bell
inequality requires that they stay the same. In effect, the Bell
inequality only holds on the common part of the four different
ensembles $\Lambda_{AC'}$, $\Lambda_{AD'}$, $\Lambda_{BC'}$, and
$\Lambda_{BD'}$, \emph{i.e.}, for correlations of the form
\begin{equation}
  \label{eq:13}
  E(AC'|\Lambda_{AC'}\cap\Lambda_{AD'}
  \cap\Lambda_{BC'}\cap\Lambda_{BD'}).
\end{equation}
Unfortunately our experimental data comes in the form
\begin{equation}
  \label{eq:1}
  E(AC'|\Lambda_{AC'}),
\end{equation}
so we need an estimate of the relation of the common part to its
constituents:
\begin{equation}
  \begin{split}
    \delta&=\inf_{\text{settings}} 
    \frac{P(\Lambda_{AC'}\cap\Lambda_{AD'}
      \cap\Lambda_{BC'}\cap\Lambda_{BD'})}{P(\Lambda_{AC'})}\\
    &=\inf_{\text{settings}}
    P(\Lambda_{AD'}\cap\Lambda_{BC'}\cap\Lambda_{BD'}|\Lambda_{AC'}).
  \end{split}
  \label{eq:14}
\end{equation}
This is a purely theoretical construct, not available in experimental
data, but we will relate it to experimental data below. Anyhow, fixing
this relative size, we can prove

\smallskip\noindent \emph{Theorem 2 (The CHSH inequality with
  coincidence restriction)} The prerequisites (\ref{Cp1}--\ref{Cp3})
of Theorem 1 are assumed to hold except at a null set, as is
\begin{enumerate}
\setcounter{enumi}{3}
\item \label{C2p4} \emph{Coincident events.}  Correlations are
  obtained on subsets of $\Lambda$, namely on
  \begin{equation}
    \Lambda_{AC'},\;\Lambda_{AD'},\;\Lambda_{BC'},
    \text{ or }\Lambda_{BD'}.\tag{iv}
  \end{equation}
\end{enumerate}
Then
\begin{equation}
  \Big| E(AC'|\Lambda_{AC'})+E(AD'|\Lambda_{AD'}) \Big| 
  + \Big|E(BC'|\Lambda_{BC'})-E(BD'|\Lambda_{BD'}) \Big|
    \le 4-2\delta.
\end{equation}

\noindent
\emph{Proof.} The proof consists of two steps; the first part is
similar to the proof of Theorem~1, using the intersection
\begin{equation}
  \Lambda_{\text{I}}
  =\Lambda_{AC'}\cap\Lambda_{AD'}\cap\Lambda_{BC'}\cap\Lambda_{BD'}, 
  \label{eq:intersection}
\end{equation}
on which coincidences occur for all relevant settings.  This ensemble
may be empty, but only when $\delta=0$ and then the inequality is
trivial, so $\delta>0$ can be assumed in the rest of the proof. Now
(\ref{Cp1}--\ref{Cp3}) yields
\begin{equation}
    \Big| E(AC'|\Lambda_{\text{I}})+E(AD'|\Lambda_{\text{I}}) \Big|
    +\Big| E(BC'|\Lambda_{\text{I}})-E(BD'|\Lambda_{\text{I}}) \Big|
    \le 2.
  \label{eq:11}
\end{equation}
The second step is to translate this into an expression with
$E(AC'|\Lambda_{AC'})$ and so on.  For brevity, let
$\Lambda_{\text{O}}=\Lambda_{AD'}\cap\Lambda_{BC'}\cap\Lambda_{BD'}$
and denote ``set complement'' by $\complement$. Then,
$\Lambda_{\text{I}}=\Lambda_{\text{O}}\cap\Lambda_{AC'}$ and
\begin{equation}
  \label{eq:18}
    E(AC'|\Lambda_{AC'})=
    P(\Lambda_{\text{O}}|\Lambda_{AC'})
    E(AC'|\Lambda_{\text{O}}\cap\Lambda_{AC'})
    +P(\Lambda_{\text{O}}^{\complement}|\Lambda_{AC'})
    E(AC'|\Lambda_{\text{O}}^{\complement}\cap\Lambda_{AC'}).
\end{equation}
We now have
\begin{equation}
  \begin{split}
    \Big|E&(AC'|\Lambda_{AC'})-\delta E(AC'|\Lambda_{\text{I}})\Big|\\
    &\le\Big|P(\Lambda_{\text{O}}^\complement|\Lambda_{AC'})
    E(AC'|\Lambda_{\text{O}}^\complement\cap\Lambda_{AC'})\Big|
    +\Big|P(\Lambda_{\text{O}}|\Lambda_{AC'})
    E(AC'|\Lambda_{\text{O}}\cap\Lambda_{AC'})
    -\delta E(AC'|\Lambda_{\text{I}})\Big|\\
    &\le P(\Lambda_{\text{O}}^\complement|\Lambda_{AC'})
    E\big(|AC'|\big|\Lambda_{\text{O}}^\complement
    \cap\Lambda_{AC'}\big)
    +\Big(P(\Lambda_{\text{O}}|\Lambda_{AC'})
    -\delta\Big) E\big(|AC'|\big|\Lambda_{\text{I}}\big)\\
    &\le P(\Lambda_{\text{O}}^\complement|\Lambda_{AC'})
    +P(\Lambda_{\text{O}}|\Lambda_{AC'})-\delta=1-\delta,
    \label{eest}
  \end{split}
\end{equation}
which, together with ineq.~(\ref{eq:11}) and the triangle inequality,
yields the desired result after some simple manipulations. \qed

Let us now relate this to experimental quantities. In this context,
the quantity of greatest interest is the probability of coincidence:
\begin{equation}
  \label{eq:3}
  \gamma\defeq\inf_{\text{settings}}P(\Lambda_{AC'}).
\end{equation}
We now have
\begin{equation}
  \delta\ge4-\frac3\gamma
\end{equation}
because (Bonferroni)
\begin{eqnarray}
  \label{eq:7}
  P(\Lambda_{AD'}\cap\Lambda_{BC'}\cap\Lambda_{BD'}|\Lambda_{AC'})
    \ge
    P(\Lambda_{AD'}|\Lambda_{AC'})+P(\Lambda_{BC'}|\Lambda_{AC'})
    +P(\Lambda_{BD'}|\Lambda_{AC'})-2,\quad
\end{eqnarray}
and
\begin{equation}
  \label{eq:4}
  \begin{split}
    P(\Lambda_{BD'}|\Lambda_{AC'})&=\frac{P(\Lambda_{AC'}
      \cap\Lambda_{BD'})}{P(\Lambda_{AC'})}
    =\frac{P(\Lambda_{AC'})+P(\Lambda_{BD'})
      -P(\Lambda_{AC'}\cup\Lambda_{BD'})}{P(\Lambda_{AC'})}
    \\&
    \ge 1+\frac{P(\Lambda_{BD'})-1}{P(\Lambda_{AC'})}\ge
    1+\frac{\gamma-1}{\gamma} = 2-\frac1\gamma.
  \end{split}
\end{equation}
Putting this into our modified CHSH inequality we arrive at
\begin{equation}
    \Big| E(AC'|\Lambda_{AC'})+E(AD'|\Lambda_{AD'}) \Big| 
    + \Big|E(BC'|\Lambda_{BC'})-E(BD'|\Lambda_{BD'}) \Big|
    \le\frac6\gamma-4.
\end{equation}
The bound for violation by quantum mechanics here is
$\gamma>3-\frac3{\sqrt2}\approx0.8787$, which is considerably higher
than the corresponding value for the detector-efficiency case,
$\frac1{\sqrt2}\approx0.7071$.

Let us see whether this bound is necessary and sufficient. At the same
time, we answer the question if it would be possible to lower the
bound by putting further natural constraints on the model. This will
be done by construction of an \emph{ad hoc} model that will give the
quantum predictions at the settings $a=0$, $b=\pi/2$, $c=\pi/4$, and
$d=-\pi/4$, and the additional natural constraints are: it will only
use local data, even for timing; the marginal distributions are
correct; there is full correlation if the settings are equal at the
two sites; and the coincidence probability is the same at our
specified pairs of settings.

\begin{figure}[htbp]
  \begin{center}
    \includegraphics[scale=.8]{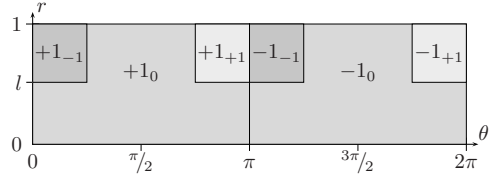}
    \caption{Outcome pattern for the detectors. The subscripts are
      the detection times. Thus $+1_0$ means outcome
      $+1$ at time $0$.}
    \label{fig:1}
  \end{center}
\end{figure}

The model is as follows: the hidden variable $\lambda$ is a pair
$(\theta,r)$ of coordinates, uniformly distributed over the rectangle
indicated in Fig.~\ref{fig:1}. The local detector setting corresponds
to a shift in the $\theta$-direction of the pattern, with wrap-around
when necessary. The result is obtained according to the diagram (the
subscript is the detection time which can be $\pm1$ or 0). To make the
behaviour interesting we choose $\Delta{}T$ to be 3/2, so that a
time-difference of zero or one time unit(s) is a coincidence while a
time-difference of two time units will not be a coincidence.

\begin{figure}[htbp]
  \begin{center}
    \includegraphics[scale=0.8]{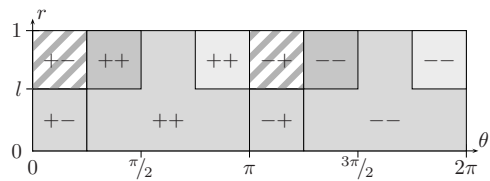}      
    \caption{Coincidences occur as follows: the events are
      truly coincident in the middle-gray areas, and since %
      $\Delta T>1$ events are ``coincident'' in the other grey areas,
      but since $\Delta T<2$ events are ``non-coincident'' in the
      hatchmarked areas.}
    \label{fig:4}
  \end{center}
\end{figure}

For example, for the settings $a=0$ and $c=\pi/4$ at the two sites,
there will be coincidences at the $\lambda$s indicated in
Fig.~\ref{fig:4}, so that the probability of coincidence is $3/4+l/4$,
while the probability of getting $++$ or $--$ is $3/4$. For the
settings $b=\pi/2$ and $d=-\pi/4$ at the two sites, the probability of
coincidence would again be $3/4+l/4$, while the probability of getting
$++$ or $--$ would only be $l/4$, so that
\begin{equation}
  E(AC'|\Lambda_{AC'})=E(AD'|\Lambda_{AD'})
  =E(BC'|\Lambda_{BC'})
  =-E(BD'|\Lambda_{BD'})=\frac{3-l}{3+l}.
\end{equation}

Setting $(3-l)/(3+l)=1/{\sqrt2}$, \emph{i.e.},
$l=3(3-2\sqrt2)\approx0.5147$ we obtain
\begin{equation}
  \label{eq:10}
  \gamma=\frac{3+l}4=\frac{3+3(3-2\sqrt2)}4
  =3-\frac3{\sqrt2},
\end{equation}
which saturates the derived coincidence probability bound. This model
does what we have asked so far, especially, it violates the Bell
inequality maximally. The model does not have constant coincidence
probability for \emph{all} angular settings but can be modified so
that it does \footnote{The relative delays in our ad hoc model vary in
  probability, but this can also be fixed, introducing some additional
  complexity.}. Furthermore, the interference pattern is not
sinusoidal, and delays are discrete, but this will be a subject of
further research (see \emph{e.g.}, ref.~\cite{jalar99a}).

We have seen that a useful Bell inequality does hold even if the
detection time is allowed to depend on the settings, if sufficiently
many events are simultaneous. That the inequality needs to be modified
when events are allowed to drop from the statistics is not surprising
but to be expected, cf.\ previous analysis in the low-efficiency case.
It is perhaps more surprising that the amount of coincidences needs to
be higher in this case than in the efficiency case. The reason for
this is that the set of coincidences $\Lambda_{AC'}$ factors in the
efficiency case, \emph{i.e.},
$\Lambda_{AC'}=\Lambda_A\cap\Lambda_{C'}$ (see ref.~\cite{jalar98a}),
while here the set cannot be factored. Thus, the present treatment is
a proper generalization of the previous results.  A major remaining
challenge is to extend the analysis to the situation when coincidence,
detection \emph{and} memory loopholes (see
refs.~\cite{gill03b,gill03a,gill03c}) are all present.

Several modern experiments are not affected by this loophole, such as
the \emph{ion trap experiment} by Rowe et al \cite{Rowe01}, because
there, all experimental runs produce coincidences (although locality
is not strictly enforced). Also, the \emph{``event-ready'' experiment}
proposed in ref.~\cite{Bell81} removes the coincidence loophole (see
also ref.~\cite{ZZHE}). Such an experiment uses a tri-partite system
and a third detector located, \emph{e.g.}, near the source as an
indicator that there is a pair being emitted. A detection there would
then be independent of the settings of the distant detectors and would
give the timing information needed to remove the coincidence loophole.
In the case of \emph{pulsed optical experiments}, one can use the
natural assumption that the setting-dependent delays described by
$T_{a,c}$ and $T'_{a,c}$ do not depend on the relation between pulse
length and pulse spacing. Then, if the pulse (\emph{e.g.}, the driving
pulse of a parametric downconverter) is short in comparison to the
pulse spacing, one can assume that any delays that occur will not
delay photons from the time-window of one pulse to the next, or at
least that this will happen only with very low probability. Now, the
driving pulse will provide a well-defined, pre-determined coincidence
window and this will remove the coincidence loophole. In both the
latter approaches a lowered efficiency may remain, but using an
event-ready experiment or a pulsed source (with the above two natural
assumptions) will enable use of the previous lower bound, \emph{e.g.},
from ref.~\cite{jalar98a}.

In conclusion, we have shown that the coincidence loophole can be
significantly more damaging than the well-studied detection problem.
Fortunately, the damage can be quantified and in some cases, repaired.
The results underline the importance of eliminating post-selection in
future experiments.

\acknowledgements 

J.-\AA.~L.\ was financially supported by the Swedish research council.
R.~D.~G.\ was partially funded by project RESQ (IST-2001-37559) of the
IST-FET programme of the European Union.


\newcommand{\sortnoop}[1]{}

\end{document}